\documentclass[prb,twocolumn,showpacs,preprintnumbers,amsmath,amssymb,floatfix]{revtex4}

\usepackage{graphicx}
\usepackage{dcolumn}
\usepackage{bm}
\usepackage{booktabs}
\usepackage{color}

\newcommand{\laco}{La$_{1-x}A_x$CoO$_3$}
\newcommand{\lcco}{La$_{1-x}$Ca$_x$CoO$_3$}
\newcommand{\leco}{La$_{1-x}$Eu$_x$CoO$_3$}
\newcommand{\lsco}{La$_{1-x}$Sr$_x$CoO$_3$}
\newcommand{\lbco}{La$_{1-x}$Ba$_x$CoO$_3$}
\newcommand{\lco}{LaCoO$_3$}
\newcommand{\Cod}{Co$^{3+}$}
\newcommand{\Cov}{Co$^{4+}$}
\newcommand{\Tc}{\ensuremath{T_{\rm{c}}}}
\newcommand{\aLTM}{\ensuremath{T^{\alpha}_{\rm{LTM}}}}
\newcommand{\aHTM}{\ensuremath{T^{\alpha}_{\rm{HTM}}}}
\newcommand{\aki}{\ensuremath{T^{\alpha}_{\rm{kink}}}}
\newcommand{\cpLT}{\ensuremath{T^{c_{p}}_{\rm{LT}}}}
\newcommand{\cpHT}{\ensuremath{T^{c_{p}}_{\rm{HT}}}}
\newcommand{\xcool}{\ensuremath{T^{\rm{x-ray}}_{\rm{cool}}}}
\newcommand{\xwarm}{\ensuremath{T^{\rm{x-ray}}_{\rm{warm}}}}
\newcommand{\Mag}{$M_{4\rm{K}}$}
\newcommand{\cpT}{\ensuremath{c_{\rm p}/T}}
\newcommand{\cp}{\ensuremath{c_{\rm p}}}

\begin{document}

\title{Magnetic and structural transitions in La$_{\bm{1-x}}\bm{A}_{\bm{x}}$CoO$_{\bm{3}}$ ($\bm{A}$\,=\,Ca, Sr, and Ba)}

\author{M.~Kriener$^{1,2}$, M.~Braden$^{2}$, H.~Kierspel$^{2}$, D.~Senff$^{2}$, O.~Zabara$^{2}$, C.~Zobel$^{2}$, and T.~Lorenz$^{2}$}

\affiliation{$^{1}$Department of Physics, Graduate School of Science, Kyoto University, Kyoto 606-8502, Japan\\
$^{2}$II.\ Physikalisches Institut, Universit\"{a}t zu K\"{o}ln, Z\"{u}lpicher Str. 77, 50937 K\"{o}ln, Germany}

\date{\today}

\begin{abstract}
We report thermal-expansion, lattice-constant, and specific-heat data of the series \laco\ for $0\leq x \leq 0.30$ with $A$\,=\,Ca, Sr, and Ba. For the undoped compound \lco\ the thermal-expansion coefficient $\alpha(T)$ exhibits a pronounced maximum around $T=50$\,K caused by a temperature-driven spin-state transition from a low-spin state of the Co$^{3+}$ ions at low towards a higher spin state at higher temperatures. The partial substitution of the La$^{3+}$ ions by divalent Ca$^{2+}$, Sr$^{2+}$, or Ba$^{2+}$ ions causes drastic changes in the macroscopic properties of LaCoO$_{3}$. The large maximum in $\alpha(T)$ is suppressed and completely vanishes for $x\gtrsim 0.125$. For $A$\,=\,Ca three different anomalies develop in $\alpha(T)$ with further increasing $x$, which are visible in specific-heat data as well. Together with temperature-dependent x-ray data we identify several phase transitions as a function of the doping concentration $x$ and temperature. From these data we propose an extended phase diagram for \lcco.
\end{abstract}



\maketitle

\section{Introduction}
Transition-metal oxides of the form $RM$O$_3$ containing Lanthanides $R$ and $3d$ elements $M$ attract a lot of interest because of their rich variety of physical properties. This is due to strong correlation effects among the $3d$ electrons and their hybridization with the oxygen $2p$ orbitals, which leads to a complex interplay of different degrees of freedom like orbital, electric, magnetic, and structural features resulting in complex phase diagrams. For $M$\,=\,Co an additional degree of freedom exists: Cobalt ions exhibit the possibility of spin-state transitions leading to an even more complex behavior. A prominent example is \lco, which is in the focus of research since the early 1950's.\cite{jonker53a,senaris95a,saitoh97b,asai98a,tokura98a,yamaguchi97a,kobayashi00b} Here, the Co$^{3+}$ ions feature a $3d^6$ configuration which in principle can occur in three different spin states: a low-spin (LS) ($t_{2g}^{6}e_{g}^{0}$, $S=0$), an intermediate-spin (IS) ($t_{2g}^{5}e_{g}^{1}$, $S=1$), and a high-spin (HS) state ($t_{2g}^{4}e_{g}^{2}$, $S=2$). \lco\ exhibits a non-magnetic ground state at low temperatures which is attributed to a LS state of the \Cod\ ions. With increasing temperature a magnetic moment develops above approximately 30\,K whereas the insulating behavior is preserved leading to a paramagnetic semiconductor at room temperature. Around 500\,K a metal-insulator transition is reported. The development of a magnetic moment is discussed in terms of a spin-state transition, i.\,e., a temperature-driven electron transfer from the $t_{2g}$ to the $e_g$ orbitals realizing one of the two possible higher spin states. However, the nature of the populated spin state is controversially discussed. Earlier publications suggest a LS\,--\,HS transition\cite{raccah67a,asai94a,itoh94a,senaris95a,yamaguchi96a} whereas in the 1990s the possibility of a LS\,--\,IS transition entered the discussion.\cite{potze95a,korotin96a,saitoh97b,asai98a,yamaguchi97a,kobayashi00b,zobel02a,radaelli02a} Recent spectroscopic studies turn the discussion back to a LS\,--\,HS scenario.\cite{noguchi02a,haverkort06a}

The physical properties of \lco\ can be changed drastically by the heterovalent substitution of the La$^{3+}$ ions by divalent earth-alkaline elements like $A$\,=\,Ca$^{2+}$, Sr$^{2+}$, or Ba$^{2+}$.\cite{itoh94a,sathe96a,muta02a,tsubouchi03a,kriener04a,kriener07a} This charge-carrier doping creates formally \Cov\ ions: La$_{1-x}^{3+}A_x^{2+}$Co$_{1-x}^{3+}$Co$_{x}^{4+}$O$_3^{2-}$. The \Cov\ ions feature a $3d^5$ configuration which is magnetic in all possible spin states due to the odd number of electrons: $t_{2g}^{5}e_{g}^{0}$, $S=1/2$ (LS), $t_{2g}^{4}e_{g}^{1}$, $S=3/2$ (IS), and $t_{2g}^{3}e_{g}^{2}$, $S=5/2$ (HS). Therefore, one can expect (i) a change in the electrical properties due to the hole doping, (ii) a modification of the magnetic properties, and (iii) structural changes because of the different ionic radii of the substitutents as well as of the different Co-oxidation states.

The electric and magnetic phase diagrams of the series \laco\ confirm the expectations (i) and (ii):\cite{itoh94a,kriener04a} The heterovalent substitution of La by Ca, Sr, or Ba suppresses the prominent maximum in $\chi$ which is caused by the aforementioned spin-state transition. The non-magnetic ground state vanishes very quickly with increasing doping concentration $x$. For $A$\,=\,Sr and Ba the system runs into a spin-glass phase for intermediate $x$ and orders ferromagnetically at higher doping concentrations $x\geq 0.2$.\cite{itoh94a,anilkumar98a,masuda03a,hoch04a,kriener04a} In some studies the ferromagnetic phase is attributed to a cluster-glass phase, but with our techniques we are not able to distinguish between them. For simplicity, we will refer to the magnetic phase as a ferromagnetic phase in the following, but one has to keep in mind that the magnetic ordering may be more complex. Simultaneously with the occurrence of ferromagnetic order, the Sr- and Ba-doping series become metallic, too.

In a previous study we did not observe indications of a spin-glass phase in \lcco,\cite{kriener04a} in contrast to some other studies reporting the presence of a frustrated magnetic phase with spin-glass-like behavior.\cite{baily02a,muta02a,szymczak05a,phelan07a} Compared to the Sr- and Ba-doping series our magnetization data is qualitatively different in the case of $A$\,=\,Ca. Instead of a low magnetic moment \Mag\ (= magnetization at 4\,K) observed for the Sr- and Ba-doping series, the Ca-doped samples exhibit clear signatures of ferromagnetic order already for small doping concentrations with \Mag\ values of the same order of magnitude as the values found for the highly Sr- and Ba-doped ferromagnetic compounds. Nevertheless, the saturation moment for Ca doping at larger $x$ is somewhat smaller than for $A$\,=\,Sr and Ba. Phenomenologically, the electric and magnetic properties can be understood in a double-exchange model as described in detail in Ref.\,\onlinecite{kriener04a}.

The substitution of La by Sr does not change the rhombohedral R$\bar{3}$c symmetry of the crystal structure at least up to $x\approx 0.5$, see e.\,g.\ Refs.\,\onlinecite{sathe96a}, \onlinecite{kriener04a}, \onlinecite{senaris95b,ganguly94a,phelan07a,caciuffo99a}. For Ba doping most of the available reports agree that the system finally realizes a cubic structure\cite{troyanchuk98a,ganguly99a,fauth01a,mandal04a,phelan07a} although there is one report of an orthorhombic description for $x=0.5$.\cite{moritomo98a} However, the critical concentrations and/or critical temperatures found in the various reports are rather different. These discrepancies mainly concern high Ba concentrations ($x\gtrsim 0.4$), while in the doping and temperature range of the present study ($0\leq x \leq 0.3$; $T<300$\,K) \lbco\ is found to be rhombohedral.\cite{ganguly99a,kriener04a,mandal04a,phelan07a} For A\,=\,Ca the situation is more controversial: Several publications report a reduction of the rhombohedral distortion upon increasing Ca concentration, but it remains controversial which structure is eventually realized. Early reports treated the compositions $0.1 \leq x \leq 0.35$ as rhombohedral,\cite{zock95a,ganguly99a} in contrast to Ref.\,\onlinecite{muta02a} which suggests a superposition of rhombohedral and pseudocubic phase fractions for $0.2<x<0.5$ at room temperature. Recently, the picture became more consistent and it clearly turned out that there is a structural transition in \lcco. In 2004, a neutron scattering study by Burley {\it et al.}\cite{burley04a} reported a structural phase transition from rhombohedral (space group R$\bar{3}$c) for $x<0.15$ to orthorhombic (space group Pbnm) for $x>0.2$ with both phases coexisting in the intermediate doping range, in agreement with our own x-ray diffraction results\cite{kriener04a,kriener07a} and a more recent temperature-dependent neutron scattering study on \lcco\ by Phelan {\it et al.}\cite{phelan07a} This structural phase transition occurs in \leco, too,\cite{baier05a} and can be traced back to the internal bond-length mismatch between La-O and Co-O bonds, which is critically enhanced when La is partially substituted by the smaller Ca or Eu ions.

In this paper, we report measurements of the thermal-expansion coefficient $\alpha$ and of the specific-heat \cp\ of \lcco, which allow a very detailed study of the doping and temperature dependence of this structural phase transition and also of the magnetic ordering. Moreover, we compare the results for \lcco\ with those for \lsco\ and \lbco. The structural ($A$\,=\,Ca) and magnetic ($A$\,=\,Ca, Sr, Ba) phase transitions cause clear anomalies in both properties. Together with temperature-dependent x-ray diffraction data we confirm the structural phase boundary of \lcco\ presented in Ref.\,\onlinecite{phelan07a}. However, we observe additional large low-temperature anomalies in the thermal-expansion and rather small ones in the specific-heat data, which suggests a new phase boundary in this phase diagram.

The paper is organized as follows: In the next section we introduce the samples used and describe the experimental setups. Then we present thermal-expansion and specific-heat data for the series with $A$\,=\,Sr and Ba. We proceed with presenting the corresponding data for $A$\,=\,Ca where the resulting phase diagram is substantially more complex. Finally, the paper is summarized.

\section{Experiment}
The samples studied in this work are from the same batches as those used in our previous studies of their electric and magnetic properties. The preparation details are given in Refs.\,\onlinecite{kriener04a} and \onlinecite{berggold05a}. Here we use samples with the following doping concentrations. $A$\,=\,Ca: $x=0.05$, 0.1, 0.15, 0.17, 0.19, 0.2, 0.21, 0.23, 0.25, 0.27, 0.3 (polycrystals), and $x=0.03$ (single crystal); $A$\,=\,Sr: $x=0$, 0.002, 0.01, 0.02, 0.04, 0.08, 0.125, 0.15, 0.18, 0.25, and 0.3 (single crystals); $A$\,=\,Ba: $x=0.05$, 0.1, 0.15, 0.2, 0.25 (polycrystals), and $x=0.1$ (single crystal).

High-resolution measurements of the linear thermal-expansion coefficient $\alpha=1/L\cdot\partial L/\partial T$ were performed using a home-built capacitance dilatometer in the temperature range from 4.2\,K to $\sim 180$\,K.\cite{pott83a} Above 180\,K this dilatometer suffers from small irreproducible effects, which can prohibit reliable measurements in the high-temperature range unless the expansion effects caused by the sample are rather large. This is the case for \lcco\ and, therefore, the temperature range could be extended up to room temperature for this doping series. Specific-heat measurements were carried out using a home-built calorimeter based on a ''continuous-heating'' method \cite{KierspelDok} in the temperature range from $\sim 25$\,K to 300\,K and using a commercial calorimeter working with a ''relaxation-time'' method in the temperature range from 1.8\,K to 330\,K (Quantum Design, PPMS). Structural data were measured by x-ray diffraction on a Siemens D5000 diffractometer upon cooling and warming in the temperature range between 15\,K and room temperature using a home-built cryostat. The analysis of this data was carried out by applying the Rietveld technique using the program \textsc{Fullprof}.\cite{Fullprof}

\section{L\lowercase{a}\boldmath$_{1-x}A_{x}$C\lowercase{o}O$_{3}$ with $A=$\,S\lowercase{r}, B\lowercase{a}}

\subsection{Thermal Expansion}

\begin{figure}
\centering
\includegraphics[width=8.5cm,clip]{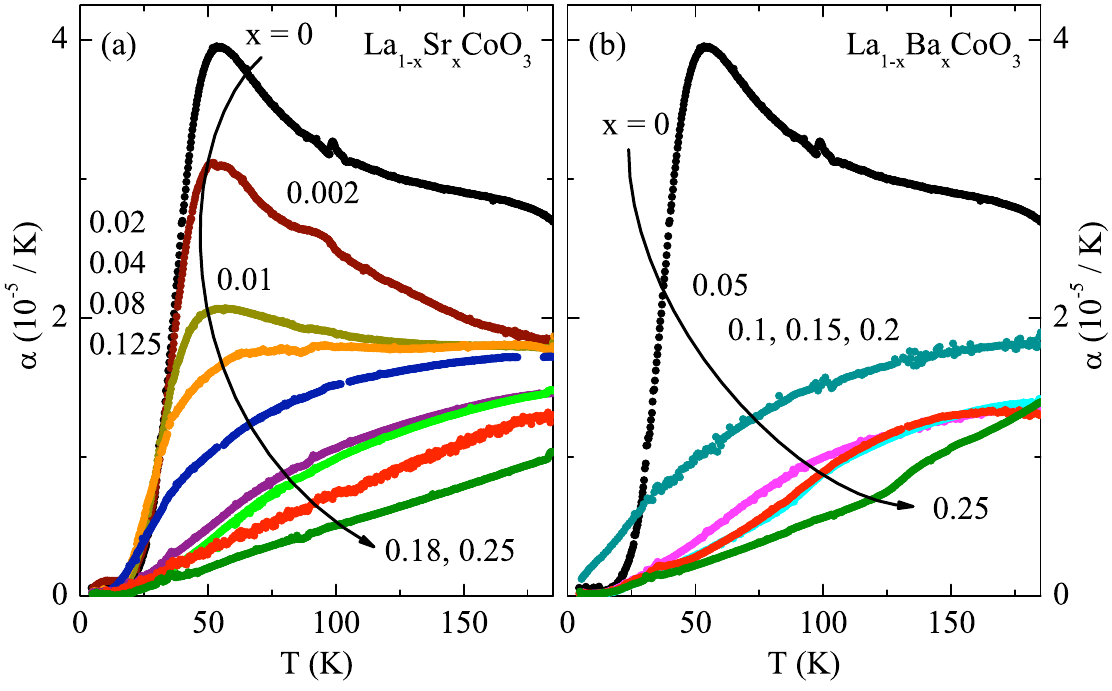}
\caption[]{(color online). Thermal-expansion coefficient $\alpha$ vs.\ $T$ of (a) \lsco\ and (b) \lbco. The arrows signal the direction of increasing $x$. The curves for $x=0.08$ and 0.125 in panel (a) and for $x=0.15$ and 0.2 in panel (b), respectively, lie almost on top of each other.}
\label{lscolbcoalpha}
\end{figure}

Fig.\,\ref{lscolbcoalpha} displays the thermal-expansion coefficients $\alpha(T)$ of (a) the \lsco\ and (b) the \lbco\ series. It is remarkable that nearly all samples exhibit a very small or even negative thermal expansion at low temperatures. This behavior is most likely related to the low-frequency rotation-phonon modes of the CoO$_6$ octahedron, whose increasing thermal population frequently causes low or negative thermal expansion. The pronounced anomaly above approximately 25\,K which broadens with $x$ is due to the spin-state transition of the \Cod\ ions as discussed in detail in Refs.\,\onlinecite{zobel02a} and \onlinecite{baier05a}.

\textbf{\textit{A}\,=\,Sr:} Already a small doping concentration such as $x=0.002$ strongly affects the spin-state transition causing the maximum in $\alpha(T)$ of the undoped compound. The rapid suppression of the $\alpha(T)$ maxima agrees well with the magnetization data.\cite{kriener04a} However, in $\alpha(T)$ a characteristic maximum or shoulder remains observable up to $x=0.125$ whereas the strong increase of the magnetization prevents a detailed analysis of the spin-state transition already for $x\gtrsim 0.01$. Thus, the thermal expansion is a much more sensitive probe for detecting spin-state transitions than other thermodynamic quantities. The onset of the maximum/shoulder around 25\,K hardly changes with increasing $x$, but its shape continuously flattens. Nevertheless, it is clearly present up to $x=0.125$, meaning that even for this large doping concentration there is still a sizeable amount of \Cod\ ions contributing to the spin-state transition. The spin-glass freezing realized in the Sr series for $0.04 < x \lesssim 0.18$ apparently does not affect the thermal-expansion coefficient. For $x\gtrsim 0.18$, the doping concentration above which magnetic order is established, a linear temperature dependence of $\alpha$ is observed. The magnetic transitions cannot be seen in our data because the corresponding \Tc\ values are above 180\,K.

\textbf{\textit{A}\,=\,Ba:} The overall behavior of \lbco\ is similar to that of the Sr-doped series, except for the 5\,\%-Ba-doped sample, where the exponential drop of $\alpha(T\rightarrow 0\,{\rm K})$ is missing. This result is consistent with the non-monotonic behavior of our magnetization data\cite{kriener04a} of this series. The magnetic moment \Mag\ at 4\,K for $x=0.05$ Ba is higher than for the samples in the intermediate doping range $0.1\leq x \leq 0.15$ which all exhibit a spin-glass phase with an $x$-independent value of \Mag. The non-monotonic behavior of $M(T)$ and $\alpha(T)$ have probably the same origin: Compared to Sr$^{2+}$ and La$^{3+}$, the Ba$^{2+}$ ion is much bigger and causes therefore stronger local disorder in \lco\ (compare the discussion in Ref.\,\onlinecite{kriener04a}). Obviously this leads to a more efficient suppression of the spin-state transition which is responsible for the exponential drop of $\alpha(T)$ upon cooling. For $0.1 \leq x \leq 0.2$ only very broad shoulders of $\alpha(T)$ are present. Thus, all curves of this Ba-doping range resemble $\alpha(T)$ of the $x=0.125$ Sr-doped crystal. However, we observe some kinks in $\alpha(T)$, which become more pronounced with increasing $x$. A kink in the same temperature interval is also present in the magnetization data of the $x=0.15$ Ba-doped sample,\cite{kriener04a} but we did not find corresponding kinks in the data for $x=0.2$ and 0.25. The kinks are not related to the onset of magnetic order, which takes place at higher temperatures, i.\,e., $T_c\gtrsim 180$\,K. Therefore, the origin of these kinks remains unclear.

\subsection{Specific Heat}
\begin{figure}
\centering
\includegraphics[width=8.5cm,clip]{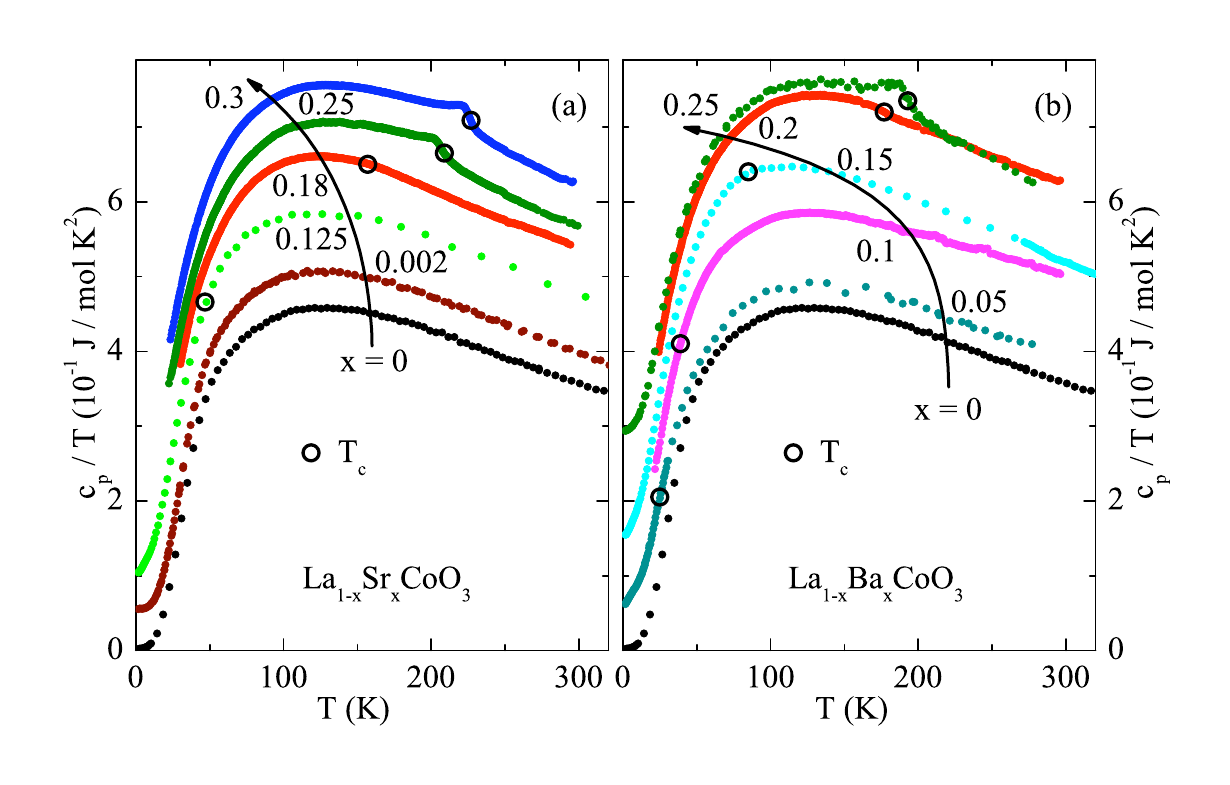}
\caption[]{(color online). Specific Heat displayed as \cpT\ vs. $T$ of (a) \lsco\ and (b) \lbco. For clarity the data for different $x$ are shifted by $+0.05$\,J/mol K$^2$ with respect to each other. The black circles denote either the spin-glass freezing temperatures or the onset temperatures of ferromagnetic order given in Ref.\,\onlinecite{kriener04a}. The arrows signal the direction of increasing $x$.}
\label{lscolbcocp}
\end{figure}
In Fig.\,\ref{lscolbcocp} the specific-heat data for \lsco\ (panel (a)) and \lbco\ (panel (b)) are displayed as \cpT\ vs $T$. Our data of \lco\ quantitatively agree with those of Ref.\,\onlinecite{stolen98a}. The black circles denote either the spin-glass freezing temperatures or the onset temperatures of ferromagnetic order for finite $x$, see Ref.\,\onlinecite{kriener04a}.

\textbf{\textit{A}\,=\,Sr:} The aforementioned spin-glass phase for a Sr content $0.04 < x \lesssim 0.18$ causes a prominent maximum in the magnetization data, but the spin-glass freezing does not cause a measurable anomaly in the specific heat. Even for the crystal with $x=0.18$ no sudden entropy release is observed, although it is located at the boundary between insulating spin-glass behavior and metallic ferromagnetism. This changes for the metallic ferromagnets with $x=0.25$ and 0.3. Here, we observe clear anomalies at temperatures which match the transition temperatures deduced from magnetization measurements as denoted by the black circles in Fig.\,\ref{lscolbcocp}\,(a). Our data confirm the observations which have been reported in Ref.\,\onlinecite{tsubouchi03a}.

\textbf{\textit{A}\,=\,Ba:} In the Ba-doped series the spin-glass freezing does not cause any measurable anomaly in the specific-heat data, either. In the ferromagnetic phase ($x=0.25$) a clear anomaly occurs at the transition temperature deduced from the magnetization measurement. However, for $x=0.2$ only a small kink appears next to the magnetic transition indicated by the black circle in Fig.\,\ref{lscolbcocp}\,(b).

In both series, $A$\,=\,Sr and Ba, the above observations indicate that the compounds do not exhibit a simple ferromagnetic order in the entire concentration range. As suggested earlier,\cite{nam00a,wu03a,kriener04a} the glass-like freezing of the magnetic moments for $x\lesssim 0.2$ is most likely caused by a competition of ferromagnetic and antiferromagnetic exchange interactions, which can prevent a spontaneous symmetry breaking at a well-defined temperature. Hence, the entropy continuously decreases upon cooling. The existence of such a frustrated spin-glass phase may also affect the ferromagnetic order (or cluster-glass phase) present for larger $x$. Probably, it is the reason for the absence of a clear anomaly in \cp\ of the compounds which are located close to the boundary between spin-glass behavior and ferromagnetic order and also explains the rather broad transitions observed for the higher doped compounds.

\section{L\lowercase{a}\boldmath$_{1-x}$C\lowercase{a}$_{x}$C\lowercase{o}O$_{3}$}
\begin{figure}[t]
\centering
\includegraphics[width=6.8cm,clip]{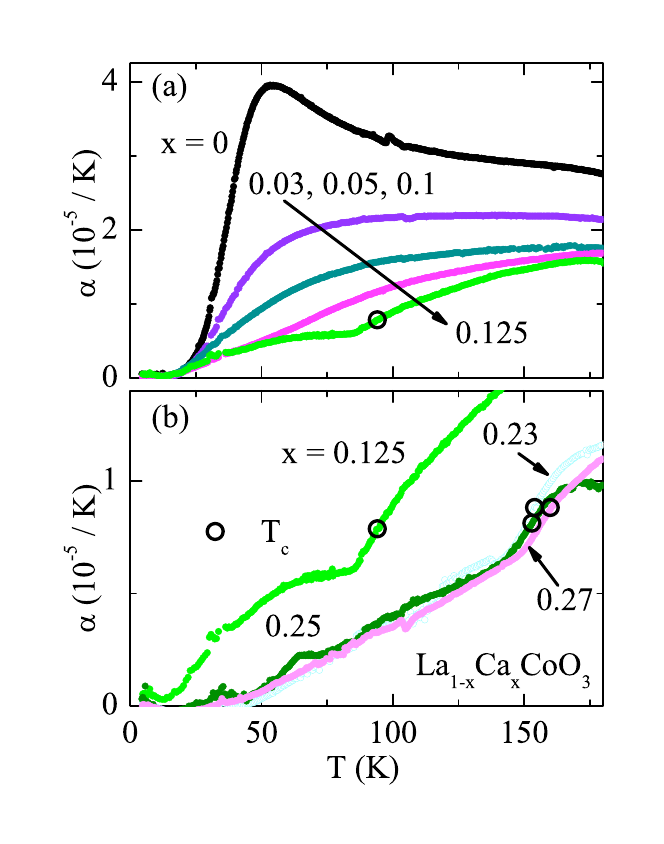}
\caption[]{(color online). Thermal-expansion coefficient $\alpha$ vs.\ $T$ of \lcco: In panel (a) $\alpha(T)$ for the low-doping region $x\leq 0.125$ and in panel (b) for the high-doping region $0.23 \leq x \leq 0.27$ is given. For comparison the data set for $x=0.125$ is shown in both panels. The ferromagnetic transition temperatures from Ref.\,\onlinecite{kriener04a} are denoted by black circles. The arrow in panel (a) signals the direction of increasing $x$.}
\label{lccoalpha}
\end{figure}
\begin{figure}
\centering
\includegraphics[width=7cm,clip]{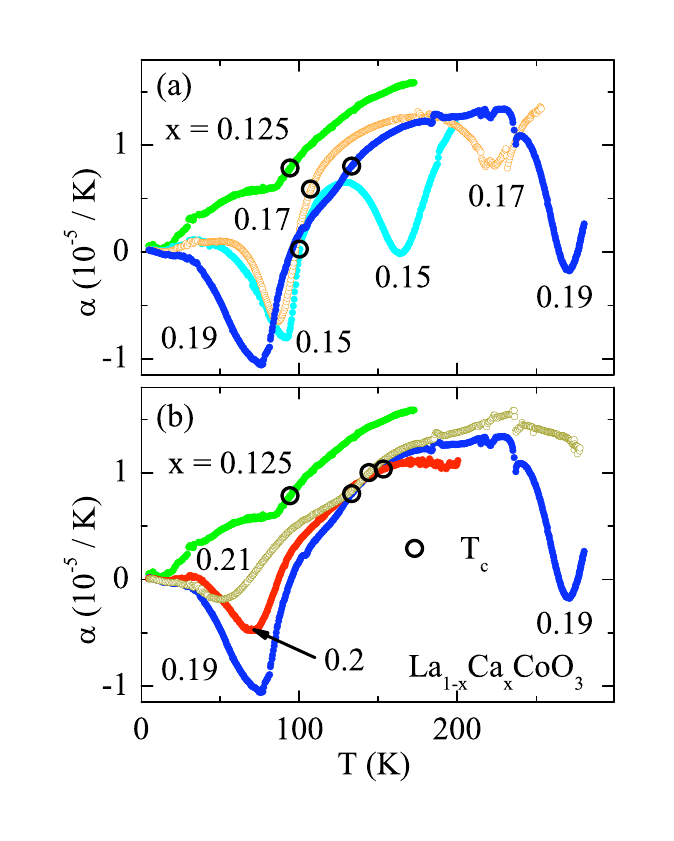}
\caption[]{(color online). Thermal-expansion coefficient $\alpha$ vs.\ $T$ of \lcco: In panel (a) $\alpha(T)$ for $0.125 \leq x\leq 0.19$ and in panel (b) for $0.19 \leq x \leq 0.21$ is given. For comparison the data sets for $x=0.125$ and 0.19 are shown in both panels. The ferromagnetic transition temperatures from Ref.\,\onlinecite{kriener04a} are denoted by black circles.}
\label{lccoalpha2}
\end{figure}

The Ca substitution enhances the intrinsic bond-length mismatch responsible for the structural distortion, as its ionic radius is much smaller than that of La. In a first view this should lead to an increase of the rotation angle of the CoO$_6$ octahedron. However, in \lcco\ an additional change in the crystal symmetry is induced. In the rhombohedral R$\bar{3}$c phase the rotation of the CoO$_6$ octahedron occurs around the [111] direction of the original cubic lattice of the perovskite. In the orthorhombic phase, space group Pbnm, there is a combination of a rotation around a cubic [001] direction and a tilting around a cubic [110] direction, see the discussion in Ref.\,\onlinecite{komarek07a}. The orthorhombic symmetry allows larger rotation and tilt distortions and apparently it may more easily accommodate internal disorder. Due to the occurrence of the structural phase transition\cite{burley04a,phelan07a} and the possible existence of an additional phase boundary (this work) the phase diagram of \lcco\ is rather complex and it is therefore discussed separately in section\,\ref{Caphadi}.

\subsection{Thermal expansion}

Figures\,\ref{lccoalpha} and \ref{lccoalpha2} show the linear thermal-expansion coefficient $\alpha$ of \lcco\ as a function of temperature. The ferromagnetic transition temperatures taken from Ref.\,\onlinecite{kriener04a} are denoted by black circles. The data for $x=0.125$ is shown in each panel of both figures for comparison. Several points are remarkable:

(i) Fig.\,\ref{lccoalpha}\,(a): Again, the pronounced maximum caused by the spin-state transition of the \Cod\ ions in the undoped compound is strongly suppressed with increasing $x$, but a shoulder remains visible up to $x\approx 0.1$.

(ii) For $x=0.125$, $\alpha(T)$ features a kink around 85\,K, which is about 10\,K below the transition temperature to ferromagnetic order (black circle). As shown in Fig.\,\ref{lccoalpha}\,(b), similar kinks are also present in $\alpha(T)$ of the highly-doped compounds with $0.23 \leq x \leq 0.27$. The kinks occur about 10\,K below \Tc, too.

(iii) Fig.\,\ref{lccoalpha2} displays $\alpha(T)$ of the intermediate doping range $0.15 \leq x \leq 0.21$. The most prominent feature of these $\alpha(T)$ curves are two pronounced minima. In addition, there are again kinks in $\alpha(T)$ for $x\geq 0.19$, which again occur about 10\,K below \Tc. Only for $x=0.2$ a larger deviation is observed.

For $x=0.15$ the two minima are close together. With increasing $x$ the low-temperature minimum (LTM) shifts continuously towards lower and the high-temperature minimum (HTM) towards higher temperature. The additional features in the $\alpha(T)$ curves above 180\,K for $0.17\leq x \leq 0.21$ arise from the aforementioned irreproducible effects of the dilatometer. Although these effects diminish the quality of the data to some extent, the pronounced minima remain clearly identifiable. For $x=0.21$, the downward curvature suggests the occurrence of a similar minimum in $\alpha(T)$ slightly above room temperature.
\begin{figure}[t]
\centering
\includegraphics[width=8cm,clip]{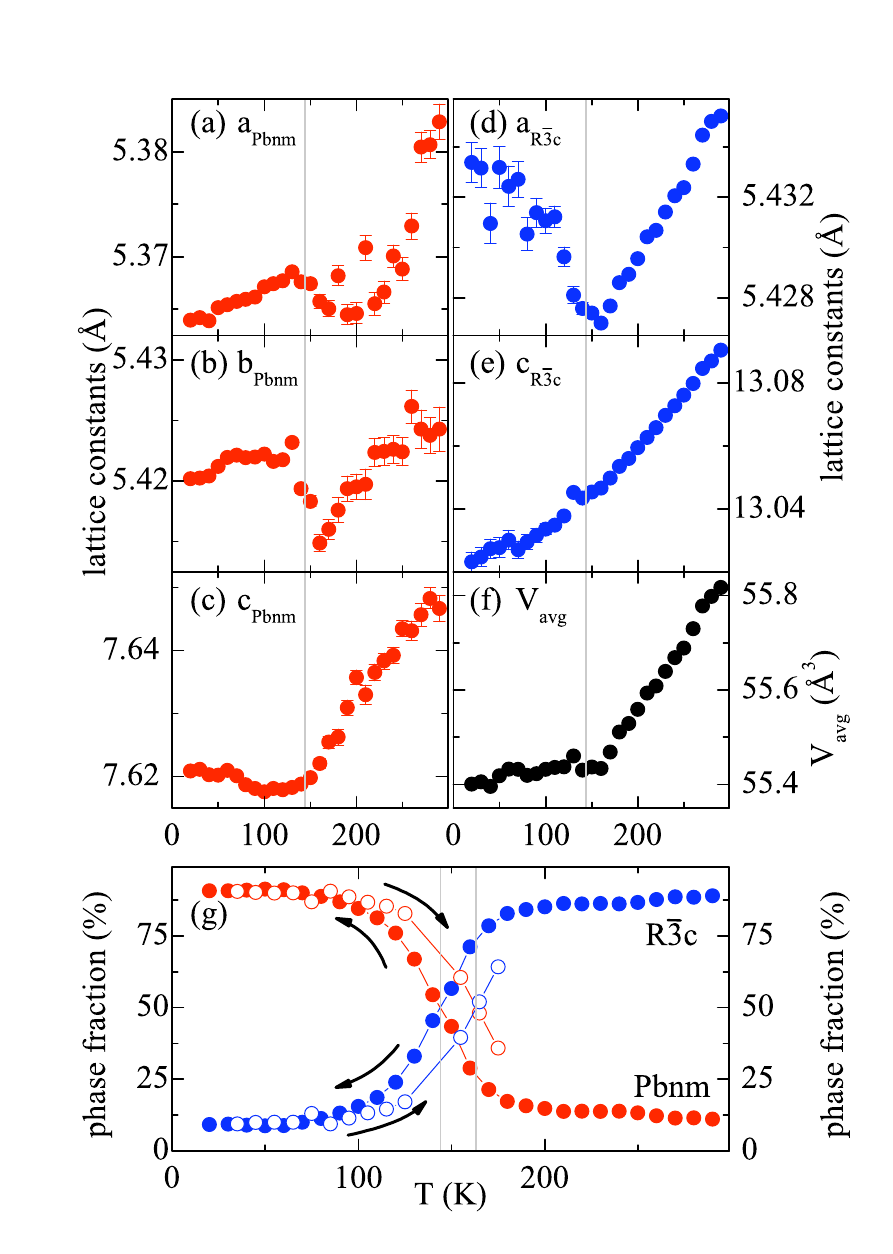}
\caption[]{(color online). Structural data of La$_{0.85}$Ca$_{0.15}$CoO$_{3}$ as a function of temperature measured upon cooling. In panels (a), (b), and (c) the orthorhombic (Pbnm) and in (d) and (e) the hexagonal (R$\bar{3}$c) lattice constants are shown. Panel (f) shows the average volume per formula unit. In (g) the orthorhombic and rhombohedral phase fractions measured with decreasing (closed symbols) and increasing (open symbols) temperature are shown. The thin grey line in panels (a)\,--\,(f) mark the temperature ($\sim 144$\,K) of equal amounts of both phases in the measurements with decreasing temperature. The second grey line in panel (g) marks the corresponding temperature ($\sim 163$\,K) in the subsequent measurement with increasing temperature.}
\label{lccoxray}
\end{figure}

As mentioned above, the room-temperature structure of \lcco\ changes as a function of $x$ from rhombohedral (space group R$\bar{3}$c, $x\leq 0.19$, throughout the paper we use the hexagonal setting of the rhombohedral lattice.) to orthorhombic (space group Pbnm, $x\geq 0.2$). Since the HTM is observed around 270\,K for $x=0.19$ and seems to shift slightly above room temperature for $x=0.21$, it appears natural that this minimum in $\alpha(T)$ signals the structural transition. This motivated us to carry out temperature-dependent x-ray diffraction studies for various concentrations $x$. Fig.\,\ref{lccoxray} shows the result for $x=0.15$. Performing two-phases fits of the diffraction patterns we find that the low-temperature majority phase of La$_{0.85}$Ca$_{0.15}$CoO$_3$ is orthorhombic and changes to rhombohedral around 150\,K. This phase transition is of first order with a pronounced hysteresis and a large temperature range of phase coexistence as can be seen in Fig.\,\ref{lccoxray}\,(g), in agreement with the results of Ref.\,\onlinecite{phelan07a}. With decreasing and increasing temperature we observe the coexistence of equal amounts of both phases at $\xcool \approx 144$\,K and $\xwarm \approx 163$\,K, respectively. The latter value agrees to the temperature \aHTM $\simeq 163$\,K of the HTM of $\alpha(T)$, which was also measured upon increasing temperature. We find a similar coincidence of $\xwarm $ and \aHTM\ for $x=0.17$ and $x=0.19$ where the structural transition is broader, see Table\,\ref{CaalphacpAnos} and Ref.\,\onlinecite{kriener07a}. Thus, we conclude that the structural phase transition is the origin of the high-temperature minimum of $\alpha(T)$.

The lattice constants given in Fig.\,\ref{lccoxray} show that the volume per formula unit is smaller in the orthorhombic phase than that in the rhombohedral phase in agreement with the idea that the smaller ionic radius on the A site is driving the transition. For a homogeneous transition one would thus expect a positive peak in the thermal expansion coefficient at the R$\bar{3}$c\,--\,Pbnm transition, whereas the high-resolution dilatometer data clearly show a negative peak. This apparent discrepancy is related to the fact that the average volume is nearly constant and is caused most likely by an anomalous expansion of the minority phase. The x-ray studies clearly indicate that the transition is not complete in both directions. Furthermore, the volume of the rhombohedral phase below the transition seems to increase even though the errors of the minority phase parameters are significantly larger. The lattice volume of the two phases is very sensitive to the spin-state distribution which is not necessarily the same in the two phases further complicating the interpretation of the structural data. The coexistance of the two phases over a large temperature intervall qualitatively agrees with neutron diffraction studies by Burley {\it et al.}\cite{burley04a} and Phelan {\it et al.}\cite{phelan07a} and point to some intrinsic inhomogeneity of \lcco. One may speculate that the large temperature range of coexisting phases might be reduced in single-crystalline samples, which could help to further clarify this issue.

The LTM of the samples with $x=0.15$ and $x=0.17$ occur rather close to the transition temperatures \Tc\ of ferromagnetic order.\cite{kriener04a} However, the characteristic temperature \aLTM\ systematically decreases with increasing $x$ while \Tc\ increases as indicated by the black circles in Fig.\,\ref{lccoalpha2}. This opposite trend rules out a relation of the LTM to the onset of ferromagnetic order. The fact that the LTM appear at the same concentration $x$ as the HTM suggests that the two types of minima may be related to each other. But again, a clear correlation appears questionable because of the opposite $x$ dependencies of \aLTM\ and \aHTM. Thus, the origin of the LTM remains unclear at present.

In all samples where the kinks in $\alpha(T)$ can be identified the corresponding temperatures \aki\ appear about 10--20\,K below \Tc, see Figs.\,\ref{lccoalpha} and \ref{lccoalpha2} and Table\,\ref{CaalphacpAnos}. This suggests some correlation between ferromagnetism and the kinks. However, for a conventional ferromagnetic order with a pressure-dependent \Tc\ one would expect a clear anomaly in $\alpha(T)$ at \Tc, similar to the corresponding anomaly of the specific heat \cp. However we do not observe clear \cp\ anomalies at \Tc, either, as discussed below. This suggests that the ferromagnetic order in \lcco\ is rather unconventional as has been proposed recently based on relaxation time measurements of the dc magnetization, too.\cite{kundu05a,szymczak05a} Indications for a spin-glass state rather than a ferromagnet have also been found in neutron diffraction for the Ca concentration $x=0.05$.\cite{phelan06a}

\subsection{Specific Heat}
\begin{figure}
\centering
\includegraphics[width=8.5cm,clip]{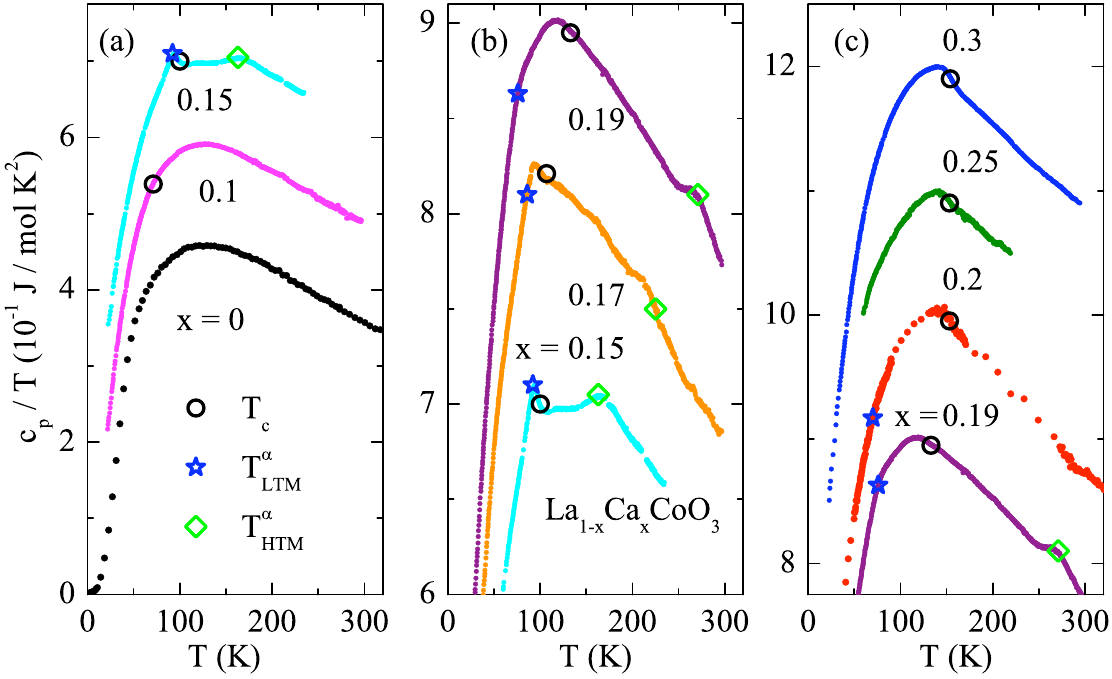}
\caption[]{(color online). Specific Heat displayed as \cpT\ vs.\ $T$ of \lcco: For clarity the data for different $x$ are shifted by $+0.1$\,J/mol K$^2$ with respect to each other. In panel (a) the curves for the doping region $0\leq x \leq 0.15$, in (b) $0.15 \leq 0.19$, and in (c) $0.19 \leq 0.3$ are given. The black circles denote the onset temperatures of ferromagnetic order from Ref.\,\onlinecite{kriener04a}. The blue stars label the LTM and the green diamonds the HTM occurring in the thermal-expansion data.}
\label{lccocp}
\end{figure}

Fig.\,\ref{lccocp} summarizes the specific-heat data for \lcco\ with $0 \leq x \leq 0.3$. Again, the black circles denote the ferromagnetic \Tc\ from magnetization data and, in addition, the temperatures \aLTM\ and \aHTM\ of the $\alpha(T)$ minima are marked by blue stars and green diamonds, respectively. For $x>0$, the \cpT\ curves also feature several anomalies: there is (i) a clear anomaly at lower temperatures \cpLT\ for $x=0.15$ and 0.17 coinciding with \Tc\ and a small slope change for $x=0.19$ not coinciding with \Tc\ resembling the behavior in $\alpha(T)$, (ii) an anomaly around \Tc\ for $x\geq 0.2$, and (iii) another anomaly at a higher temperature \cpHT, see Table\,\ref{CaalphacpAnos}. In general, all these anomalies are much less pronounced than those found in the thermal-expansion data. However, the anomalies appearing at $\cpHT$ are clearly identified and we attribute them to the structural transition since these temperatures agree well to the corresponding values \aHTM\ and \xwarm.

Concerning the magnetic-ordering temperatures, the situation is more complex. For $x=0.1$ we cannot identify an anomaly in \cpT\ around $\Tc = 70$\,K. Comparatively very large anomalies close to \Tc\ are present for $x=0.15$ and 0.17, but in both compounds \Tc\ is rather close to \aLTM\ and it is therefore not possible to allocate these anomalies in \cpT\ neither to the ferromagnetic order nor to the LTM of $\alpha(T)$. With the knowledge of the similar behavior found in $\alpha(T)$ one may argue that the ferromagnetic transition overlaps with the transition causing the LTM in $\alpha(T)$ and presumably also causes the anomalies at \cpLT\ in the specific heat. However, for $x=0.19$ these two transitions are clearly separated but none of them causes a strong \cpT\ anomaly. For $x\geq 0.2$ finally, we observe weak but clearly recognizable anomalies at the magnetic transition temperatures \Tc. Such a small entropy release at a ferromagnetic transition is rather surprising. Since the ferromagnetic single crystals of the Sr series show more pronounced anomalies of \cpT\ at \Tc, see Fig.\,\ref{lscolbcocp}, one may suspect that this is related to the polycrystalline nature of the \lcco\ samples. However, the \Tc\ anomaly of the polycrystal La$_{0.75}$Ba$_{0.25}$CoO$_{3}$ is as large as the one of the corresponding Sr-doped single crystal and much larger than those of the \lcco\ samples. This can be seen as a further indication that the nature of the magnetic ordering in the {\it entire} series \lcco\ is more complex, as has been suggested in Refs.\,\onlinecite{szymczak05a} and \onlinecite{kundu05a}, too.

\section{Phase diagram of
L\lowercase{a}\boldmath$_{1-x}$C\lowercase{a}$_{x}$C\lowercase{o}O$_{3}$}\label{Caphadi}
\begin{table*}[t]
\centering
\caption{Temperatures of the various anomalies of $\alpha(T)$ (\aLTM, \aHTM, and \aki) and $\cp(T)$ (\cpLT\ and \cpHT). For comparison, the ferromagnetic ordering temperatures (\Tc) determined by dc magnetization measurements\cite{kriener04a} are included, too. The last two columns list the transition temperatures of a first-order structural phase transition from orthorhombic to rhombohedral symmetry, deduced from temperature dependent x-ray diffraction upon cooling (\xcool) and warming (\xwarm), respectively. The columns to be compared with each other are: \aLTM with \cpLT, \aHTM with \cpHT\ and with \xwarm, and \aki\ with \Tc.
\label{CaalphacpAnos}}
\begin{tabular}{ccccccccc}
\hline
$x$     & \aLTM\ (K) & \aHTM\ (K) & \aki\ (K) & \cpLT\ (K) & \cpHT\ (K) & \Tc\ (K) & \xcool\ (K) & \xwarm\ (K)\\
\hline
$0.125$ & --   & --     & $85$  & --   & --    & 94  & --  & -- \\
$0.15$  & $92$ & $163$  & --    & $94$ & $170$ & 100 & 144 & 163\\
$0.17$  & $86$ & $225$  & --    & $93$ & $215$ & 107 & 175 & 193\\
$0.19$  & $76$ & $271$  & $125$ & $80$ & $265$ & 133 & 272 & 275\\
$0.20$  & $70$ & --     & $130$ & --   & $302$ & 153 & --  & -- \\
$0.21$  & $51$ & $>290$ & $133$ & --   & --    & 144 & --  & -- \\
$0.23$  & --   & $>300$ & $145$ & --   & --    & 154 & --  & -- \\
$0.25$  & --   & $>300$ & $145$ & --   & --    & 153 & --  & -- \\
$0.27$  & --   & $>300$ & $150$ & --   & --    & 160 & --  & -- \\
\hline
\end{tabular}
\end{table*}

\begin{figure}
\centering
\includegraphics[width=8.5cm,clip]{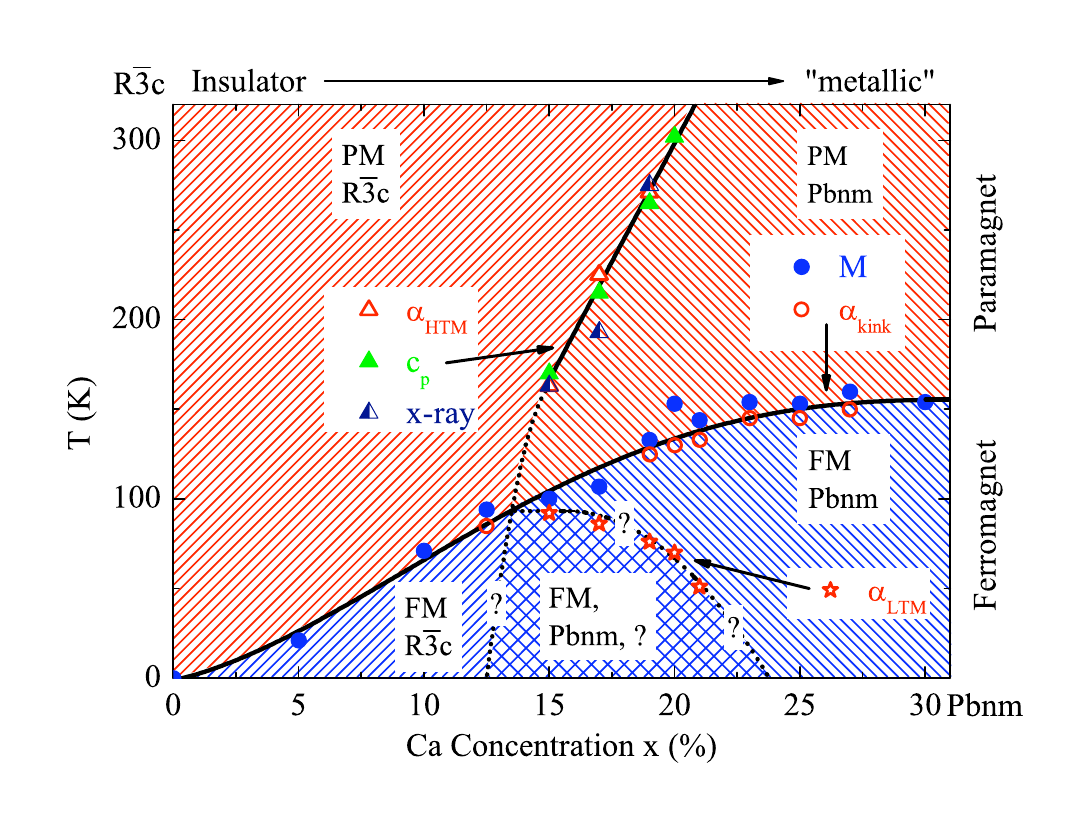}
\caption[]{(color online) Phase diagram of \lcco\ based on the
transition temperatures given in Table\,\ref{CaalphacpAnos}
obtained by magnetization (Ref.\,\onlinecite{kriener04a}),
thermal-expansion, specific-heat, and x-ray diffraction
measurements (this work). PM denotes the paramagnetic and FM the (quasi-)ferromagnetic phase, respectively. The transition from an insulator to a ''metal'' takes place only gradual in this series; compare Ref.\,\onlinecite{kriener04a}. The lines are guides to the eyes. Please note, that the structural transition is of first order. The phase boundary shown here is based on the measurements upon warming. The respective phase boundary obtained on cooling lies lower in temperature. Moreover, there is a rather large region of coexisting R$\bar{3}$c and Pbnm phases (see text).} \label{lccoPhasediagramschematic}
\end{figure}

In Table\,\ref{CaalphacpAnos}, we summarize the temperatures of all anomalies observed in the thermal-expansion (\aLTM, \aHTM, and \aki), specific-heat (\cpLT\ and \cpHT), magnetization (\Tc), and x-ray diffraction data (\xcool\ and \xwarm) of \lcco. The characteristic temperatures to be compared with each other are: \aLTM\ with \cpLT, \aHTM\ with \cpHT\ and with \xwarm, and \aki\ with \Tc. Plotting them all together yields the phase diagram in Fig.\,\ref{lccoPhasediagramschematic}. For the sake of simplicity we only show the data points determined from measurements upon increasing temperature. For the first-order structural phase transition, the corresponding phase boundary obtained on cooling is shifted to lower temperature. Both phase boundaries lie in the ''crossover'' region shown in the structural phase diagram of Ref.\,\onlinecite{phelan07a}.

(i) The magnetic phase boundary has been already reported in Ref.\,\onlinecite{kriener04a}. The magnetic phase transition is seen in the entire doping region above $x\geq 0.05$ with \Tc\ values saturating slightly above 150\,K for high Ca concentrations. The onset of magnetic order is also reflected in $\alpha(T)$ (at \aki) and in $\cp(T)$ (at \Tc). It is, however, difficult to discern a clear separation between metallic and insulating phases, the metal-insulator transition in \lcco\ is only gradual as indicated on top of the phase diagram (Fig.\,\ref{lccoPhasediagramschematic}).

(ii) Together with the information obtained by x-ray diffraction experiments another anomaly in $\alpha(T)$ (at \aHTM) and in $\cp(T)$ (at \cpHT) is identified as a first-order structural phase transition from orthorhombic to rhombohedral symmetry. For $0.15\leq x\leq 0.2 $ the corresponding transition temperature linearly increases with increasing $x$ as shown by the solid line in Fig.\,\ref{lccoPhasediagramschematic}. Note, however, that the first-order structural transition in this cross-over range is not complete. Minority phases remain visible far above and below the transition, in agreement with Ref.\,\onlinecite{phelan07a}. At higher temperatures, the slope of the structural phase boundary reduces above $x\approx 0.2$.\cite{phelan07a} A linear extrapolation to lower $x$ would end at $x\approx 0.08$ for $T\rightarrow 0$. However, the $\alpha(T)$ measurements for the samples with $x=0.1$ and $x=0.125$ do not show any indications of the HTM, which signals the structural phase transition. Thus, we conclude that the slope of this phase boundary increases and hence the phase transition vanishes in the doping region $0.125 < x < 0.15$ as denoted by the dotted line in Fig.\,\ref{lccoPhasediagramschematic}. As mentioned above the $\alpha(T)$ curves of the samples with $x\leq 0.1$ signal that in this doping range yet a sizeable fraction of the \Cod\ ions undergoes the temperature-dependent spin-state transition. Thus, one may speculate that the spin-state transition, which causes an anomalous expansion of a certain fraction of \Cod\ ions, affects the structural degree of freedom. However, a study of the related \leco\ series\cite{baier05a,EuCaionicradius} revealed that the doping-induced structural phase transition from rhombohedral to orthorhombic symmetry and the spin-state transition do not exclude each other.

(iii) The clear LTM in $\alpha(T)$ at $\aLTM$ and the rather small anomaly in $\cp(T)$ at $\cpLT$ are not visible below $x=0.125$, either, and they disappear for $x\geq 0.23$, too. The origin of this third anomaly remains unclear, although it seems that both anomalies, HTM and LTM, start at a similar doping concentration $x \gtrsim 0.125$. As already mentioned this coincides with the Ca concentration at which the signature of a temperature-driven spin-state transition of the \Cod\ ions has completely vanished. Please note that the thermal-expansion coefficient is a highly sensitive probe for this phenomenon.\cite{zobel02a,baier05a} 
Both anomalies, LTM and HTM, are of similar size in $\alpha(T)$. Hence, it is quite surprising that there are only very tiny effects (weak slope changes) in $\cp(T)$ at $\cpLT$ suggesting that this might not be a ''real'' phase transition in the thermodynamic sense. Therefore, the line corresponding to the $\aLTM$ in Fig.\,\ref{lccoPhasediagramschematic} is dotted.

\section{Summary}
In summary, we present a detailed comparative study of the thermal expansion and the specific heat of \laco\ with $A$\,=\,Ca, Sr, and Ba. The Sr- and Ba-doped series exhibit a doping-induced electrical as well as a doping- and temperature-driven magnetic transition. Our data confirm the formerly presented phase diagrams based on magnetization and resistivity measurements. Especially the onset of long-range ferromagnetic order is clearly detected in $\cp(T)$.

For \lcco\ we find up to three different anomalies (kink, LTM, and HTM in $\alpha(T)$) depending on temperature and Ca concentration. Based on these data we propose an extended phase diagram for \lcco. The comparison to the magnetization data allows to attribute the kink in $\alpha(T)$ and a much less pronounced anomaly in $\cp(T)$ to the magnetic transition. By correlating the dilatometry and x-ray data, we conclude that the HTM in $\alpha(T)$ arise from a first-order structural phase transition from a rhombohedral R$\bar{3}$c at low~$x$\,/\,low~$T$ towards an orthorhombic Pbnm symmetry at higher~$x$\,/\,higher~$T$. This structural transition, however, remains incomplete resulting in phase coexistance. The third anomaly, the LTM in $\alpha(T)$, is clearly visible in $\alpha(T)$ but surprisingly hardly affects the specific heat. Its origin remains to be clarified yet.

Finally, we comment on the question whether there are temperature-driven spin-state transitions of the \Cod\ and \Cov\ ions in \laco. Since the spin-state transition can be monitored in $\alpha(T)$ in detail, our data clearly show, that the spin-state transition observed in \lco\ is rapidly suppressed with increasing $x$. Above $x\simeq 0.125$ any indication of a temperature-driven {\em change} of the spin states has completely vanished for all three doping series \laco. From the thermal expansion data it is, however, not possible to deduce \textit{which} particular spin states are realized in the \Cod\ and \Cov\ ions, but our magnetization\cite{kriener04a} and transport\cite{berggold05a} data suggest that, at least for larger $x$, the \Cov\ ions are present in the LS and the \Cod\ ions in the IS state.

\section{Acknowledgments} This work was supported by the Deutsche Forschungsgemeinschaft through SFB 608.


\begin{thebibliography}{10}
\bibitem{EuCaionicradius}
{Assuming a ninefold coordination, Eu$^{3+}$ (1.120~\AA) and Ca$^{2+}$
  (1.180~\AA) feature both a smaller ionic radius compared to La$^{3+}$
  (1.216~\AA) which is not the case for Sr$^{2+}$ (1.310~\AA) and Ba$^{2+}$
  (1.470~\AA). The ninefold coordination is used rather than the ideal cubic
  12-fold due to the slight distortion of the systems. Radii are obtained from
  Ref.~\onlinecite{shannon76a}}.
\end{thebibliography}

\end{document}